\documentclass{appolb}
\usepackage{graphicx}
\usepackage{csquotes}
\begin{document}
\title{Measurement of the CKM angle $\phi_3$ at Belle II
\thanks{Presented at 3$^{\rm rd}$ Jagiellonian Symposium on Fundamental and Applied Subatomic Physics}%
}
\author{ P. K. Resmi \\ (on behalf of Belle II Collaboration)
\address{Indian Institute of Technology Madras, Chennai, India 600036}
}
\maketitle
\begin{abstract}
 The precise measurement of the CKM angle $\phi_3$ is important to further test the Standard Model description of $CP$ violation. The small values of the branching fractions of the decays involved in the measurement limits the precision, hence a larger dataset has to be accumulated to improve the precision. The Belle~II experiment at the SuperKEKB asymmetric-energy $e^+e^-$ collider aims to collect 50~ab$^{-1}$ of data, a factor of 50 more than that of its predecessor Belle. The accelerator has been successfully commissioned in 2016 and the first physics collisions were recorded in April 2018. The best sensitivity to $\phi_3$ can be achieved by harnessing all possible final states of $B \to D^{(*)}K^{(*)}$ decays. With the full dataset, Belle~II is expected to achieve a precision of 1$^{\circ}$ for the angle $\phi_3$. The expected sensitivities and rediscoveries from 2018 data are presented here.
\end{abstract}
  
\section{Introduction}
\label{Sec:intro}
The Cabibbo-Kobayashi-Maskawa~\cite{C,KM} unitarity triangle angle $\phi_3$ is a good probe to test the Standard Model~(SM) description of $CP$ violation. Currently, this is limited by the experimental uncertainty on $\phi_3$, which is almost a factor 10 worse than the angle $\phi_1$~\cite{HFLAV}. The $CP$-violating observables sensitive to $\phi_3$ are measured from the interference between the amplitudes of the color-favored $B^{-} \to D^{0} K^{-}$ and color-suppressed $B^{-} \to \bar{D^{0}}K^{-}$ decays, where $D$ indicates a neutral charm meson reconstructed in a final state common to both $D^{0}$ and $\bar{D^{0}}$. These are tree-level decays and the theoretical uncertainty on $\phi_3$ is $\mathcal{O}(10^{-7})$ \cite{Brod}. 

The amplitudes for the color-favored and color-suppressed decays are  $A_{\rm fav} = A$ and $A_{\rm sup} = Ar_{B}e^{i(\delta_{B} - \phi_{3})}$, respectively. Here $\delta_{B}$ is the strong-phase difference between the decay processes, and 
\begin{equation}
r_{B} = \frac{\mid A_{\rm{sup}} \mid} {\mid A_{\rm{fav}}\mid}.
\end{equation}
The statistical uncertainty on $\phi_3$ scales with $1/r_B$. The value of $r_{B}$ is approximately equal to 0.1 for $B \to DK$ decays, whereas for $B\to D\pi$, it is $0.005$. So $B \to D \pi$ decays are not very sensitive to $\phi_{3}$, but they serve as excellent control sample for $B \to DK$ to validate the signal-extraction procedure, due to the similar topology and larger sample size. The remainder of this proceedings is structured as follows: Section~\ref{Sec:methods} describes the different methods of $\phi_3$ extraction. The Belle~II experiment is described in Sec.~\ref{Sec:belle2}. The results from the data collected by the initial physics run of Belle~II is summarized in Sec.~\ref{Sec:phase2}. The expected $\phi_3$-sensitivity at Belle~II is given in Sec.~\ref{Sec:phi3}, then Sec.~\ref{Sec:sum} provides the summary.

\section{Methods for $\phi_{3}$ extraction}
\label{Sec:methods}
There are different methods to determine $\phi_3$ according to the $D$ final state under consideration. If the $D$ final state is a $CP$ eigenstate like $K^+K^-$, $\pi^+\pi^-$ or $K_{\rm S}^0\pi^0$, then the GLW~\cite{GLW} method is followed. For multibody $D$ decays like $\pi^+\pi^-\pi^0$, its $CP$-content is to be used as an external input in the measurement of $\phi_3$~\cite{MNayak}. The ADS~\cite{ADS} method is used for doubly Cabibbo-suppressed $D$ decays $K^{+}X^{-}$, where $X^{-}$ can be $\pi^{-}, \pi^{-}\pi^{0}$ or $\pi^{-}\pi^{-}\pi^{+}$. The $D$ decay parameters $r_D$ and $\delta_D$, which are the ratio of the amplitudes of the suppressed and favored $D$ decays and the $D$ strong phase, respectively, are needed as inputs. For self-conjugate multibody states such as $K_{\rm S}^{0}\pi^{+}\pi^{-}$, $K_{\rm S}^{0}K^{+}K^{-}$, $K_{\rm S}^{0}\pi^{+}\pi^{-}\pi^{0}$, the GGSZ~\cite{GGSZ} method is adopted. In this method, the $D$ phase space is divided into independent regions called \enquote{bins} and the $\phi_3$-sensitive parameters are measured from the partial rate of $B^{\pm}$ decays in each bin, which is given as
\begin{equation}
\Gamma_{i}^{\pm} \propto K_{i} + r_{B}^{2}\bar{K_{i}} +2\sqrt{K_{i}\bar{K_{i}}}(c_{i}x_{\pm} + s_{i}y_{\pm}), \label{Eg:GGSZ} 
\end{equation} 
\noindent where $x_{\pm} = r_{B}\cos (\delta_{B} \pm \phi_{3})$; $y_{\pm} = r_{B}\sin (\delta_{B} \pm \phi_{3})$. Here, $K_{i}$ and $\bar{K_{i}}$ are the fraction of flavour-tagged $D^{0}$ and $\bar{D^{0}}$ events in the $i^{\rm th}$ bin, respectively, which can be estimated from $D^{*\pm} \to D \pi^{\pm}$ decays with good precision due to their large sample size. The parameters $c_{i}$ and $s_{i}$ are the amplitude-weighted average of the cosine and sine of the strong-phase difference between $D^{0}$ and $\bar{D^{0}}$ over the $i^{\rm th}$ bin; these parameters need to be determined at a charm factory experiment like CLEO-c or BESIII, where the quantum-entangled $D^{0}\bar{D^{0}}$ pairs are produced via $e^{+}e^{-} \to \psi(3770) \to D^{0}\bar{D^{0}}$. This method allows $\phi_3$ to be determined from a single decay channel in a model-independent method.

\section{Belle~II Experiment}
\label{Sec:belle2}

The Belle~II~\cite{Belle2} detector is located at the interaction point of SuperKEKB~\cite{superkekb} asymmetric-energy $e^+e^-$ collider in Tsukuba, Japan. Belle~II is expected to accumulate a dataset corresponding to an integrated luminosity of 50~ab$^{-1}$, a factor of 50 larger than its predecessor Belle. The accelerator has been upgraded to ultimately provide a peak instantaneous luminosity 40 times more than KEKB. The Belle II detector design has also incorporated significant improvements compared to Belle. The reconstruction efficiency of $K_{\rm S}^0$ mesons is expected to improve with the larger coverage of the vertex detector. The new particle identification system is capable of providing better separation between kaons and pions. The phase~I of Belle~II happened in 2016, when accelerator commissioning took place. In 2018, the Belle~II detector, without the full vertex subsystem, was integrated at the interaction point of SuperKEKB. The collisions were recorded between 25$^{\rm th}$ April and 17$^{\rm th}$ July 2018 and this period is known as phase~II of Belle~II. A total of 472~pb$^{-1}$ of data were collected during phase~II. 

\section{Results from phase~II data}
\label{Sec:phase2}

\begin{figure}[ht!]
\begin{tabular}{cc}
\includegraphics[width=0.5\columnwidth]{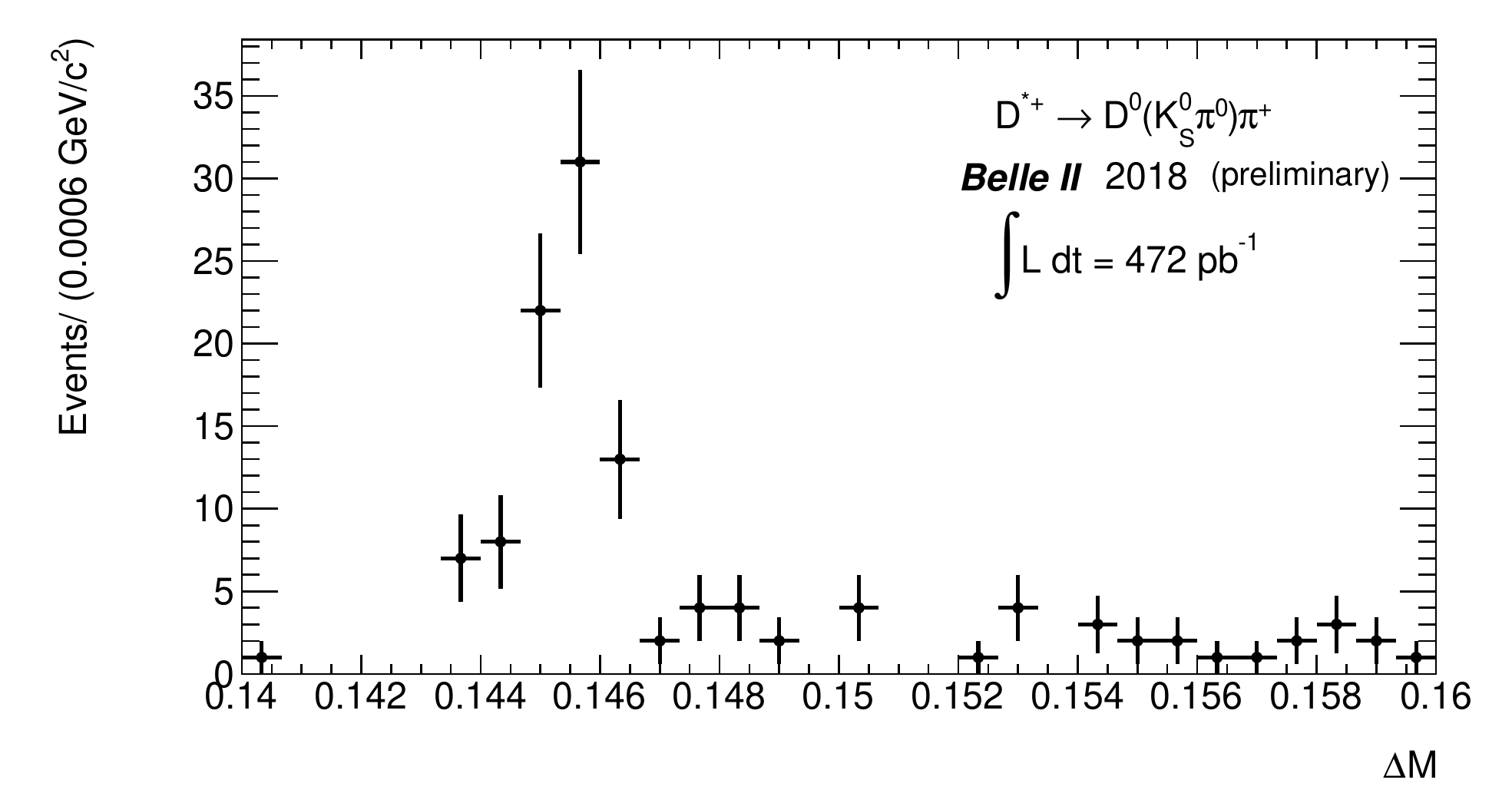} &
\includegraphics[width=0.5\columnwidth]{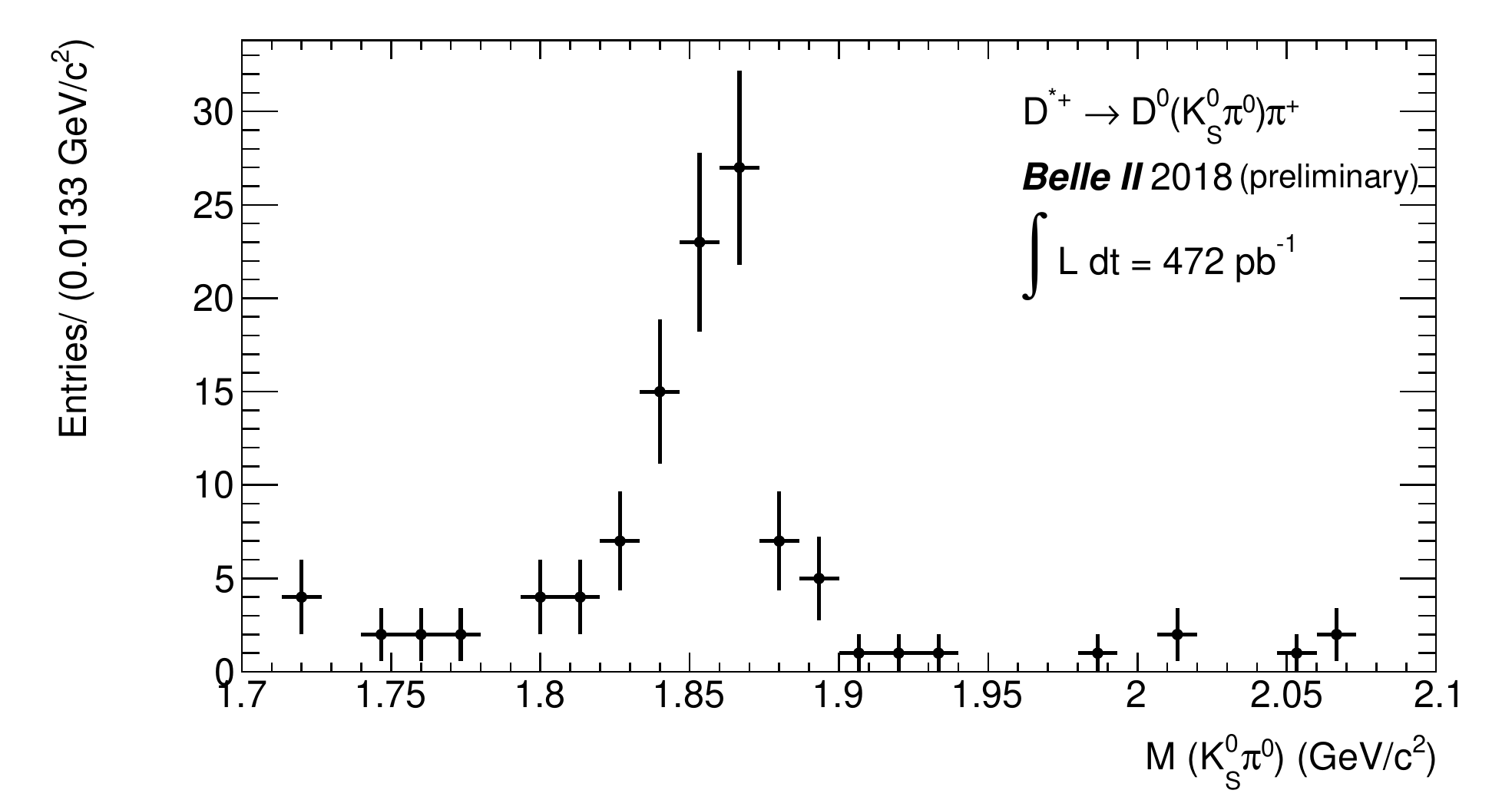} \\
\end{tabular}
\caption{$\Delta M$ (left) and $M_{D}$ (right) distributions for $D^{*\pm} \to D(K_{\rm S}^{0}\pi^{0})\pi^{\pm}_{\rm{slow}}$ decays.}\label{Fig:kspi0}
\end{figure}

\begin{figure}[ht!]
\begin{tabular}{cc}
\includegraphics[width=0.5\columnwidth]{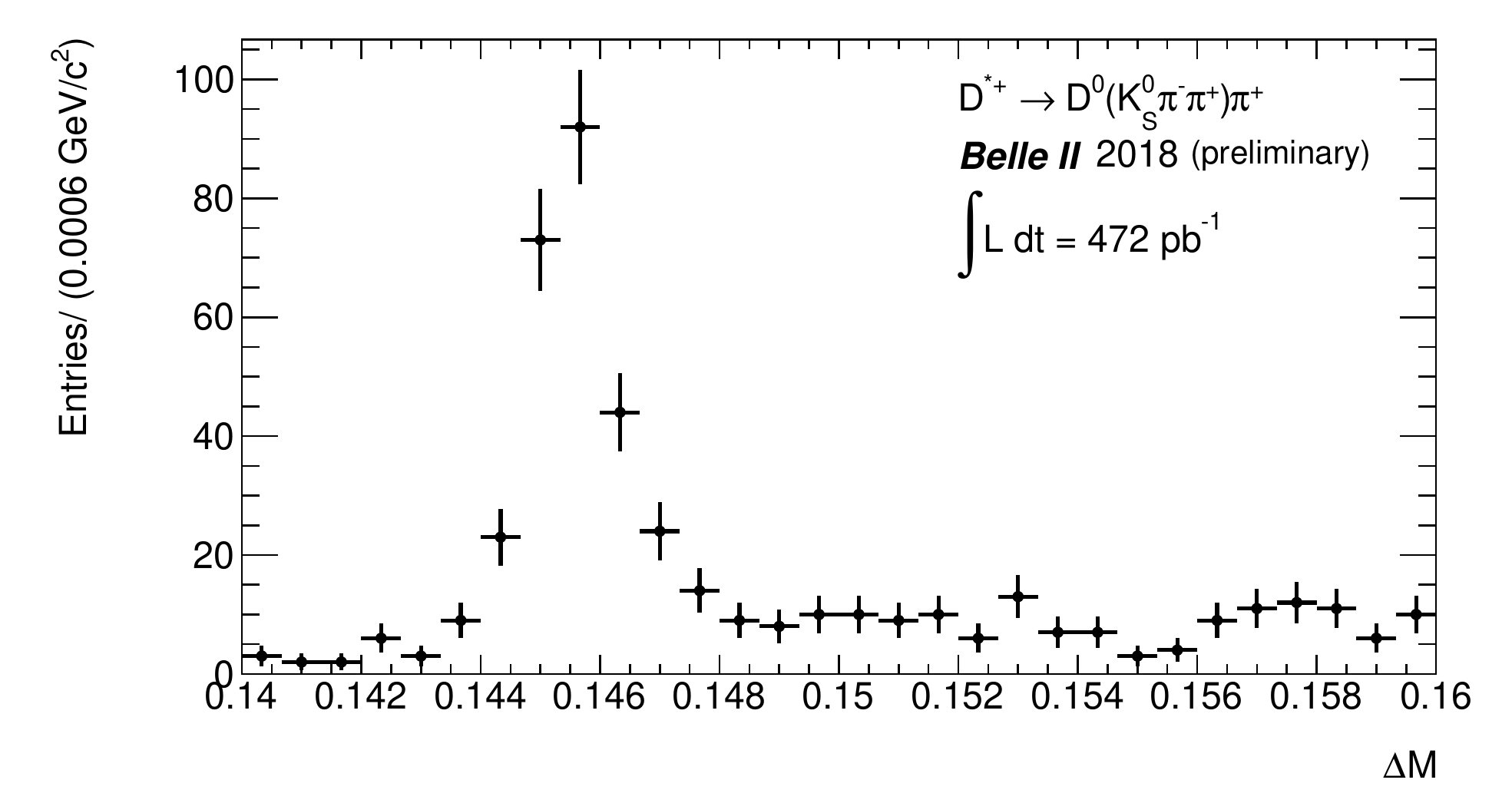} &
\includegraphics[width=0.5\columnwidth]{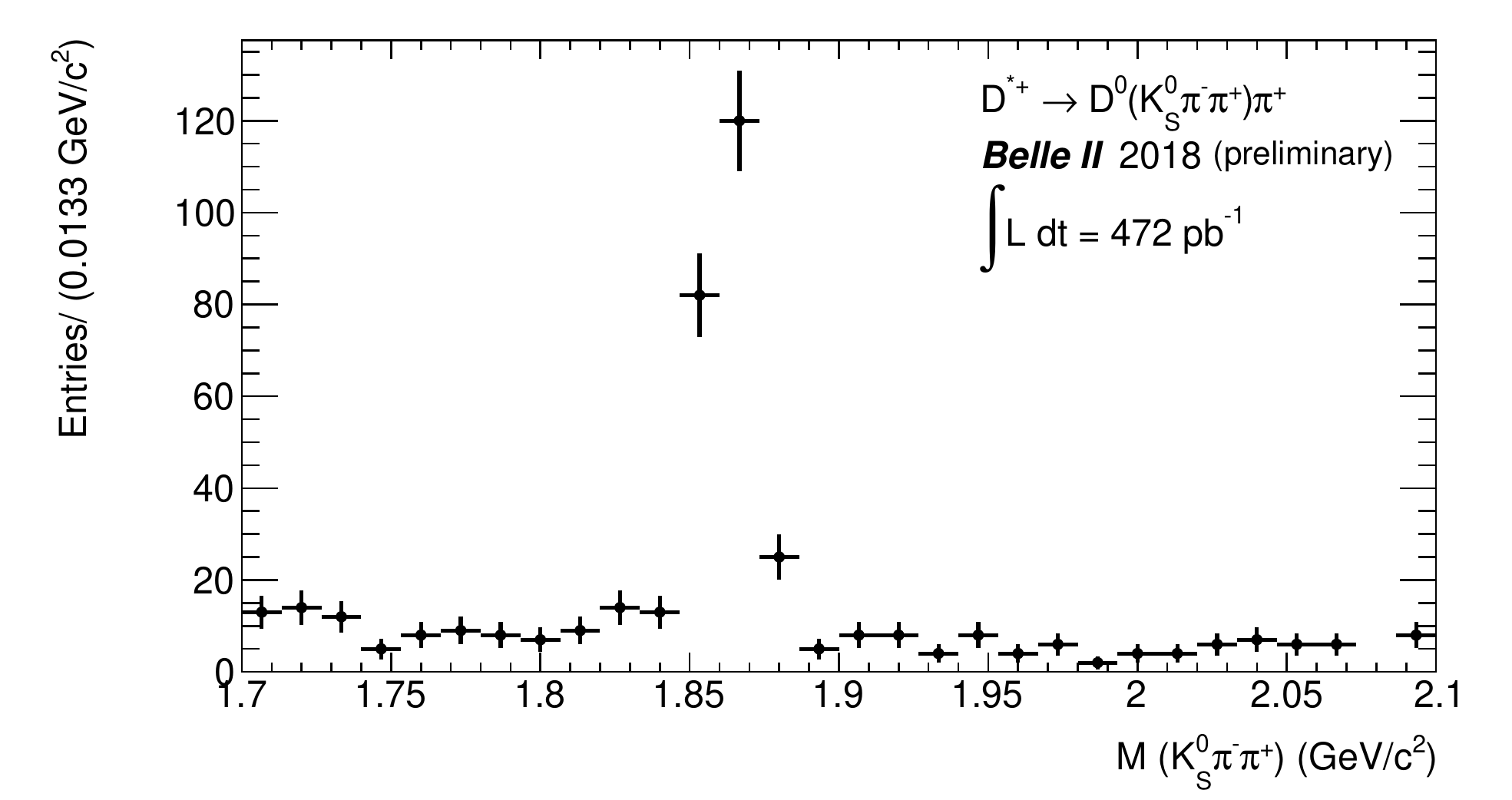} \\
\end{tabular}
\caption{$\Delta M$ (left) and $M_{D}$ (right) distributions for $D^{*\pm} \to D(K_{\rm S}^{0}\pi^{+}\pi^-)\pi^{\pm}_{\rm{slow}}$ decays.}\label{Fig:kspipi}
\end{figure}

The data from the phase~II run is helpful in assessing the performances of the accelerator and detector. A number of $D^*$ and $B$ decay modes have been rediscovered. These include various $D$ final states: $K_{\rm S}^0\pi^0$, which is a $CP$-odd eigenstate, $K^+K^-$, which is a $CP$-even as well as a singly Cabibbo-suppressed mode, and multibody self-conjugate states $K_{\rm S}^0\pi^+\pi^-$ and $K_{\rm S}^0\pi^+\pi^-\pi^0$. The observable $\Delta M$, the difference between $M_{D^*}$ and $M_D$, and $M_D$ distributions of $K_{\rm S}^0\pi^0$ and $K_{\rm S}^0\pi^+\pi^-$ candidates are shown in Figs.~\ref{Fig:kspi0} and \ref{Fig:kspipi}, respectively. The resolution is comparable to the expected values from Belle~II Monte Carlo. These analyses indicate the capability of Belle~II to reconstruct a variety of final state particles, including the neutral ones.


%\clearpage

\begin{figure}[ht!]
\begin{tabular}{cc}
\includegraphics[width=0.5\columnwidth]{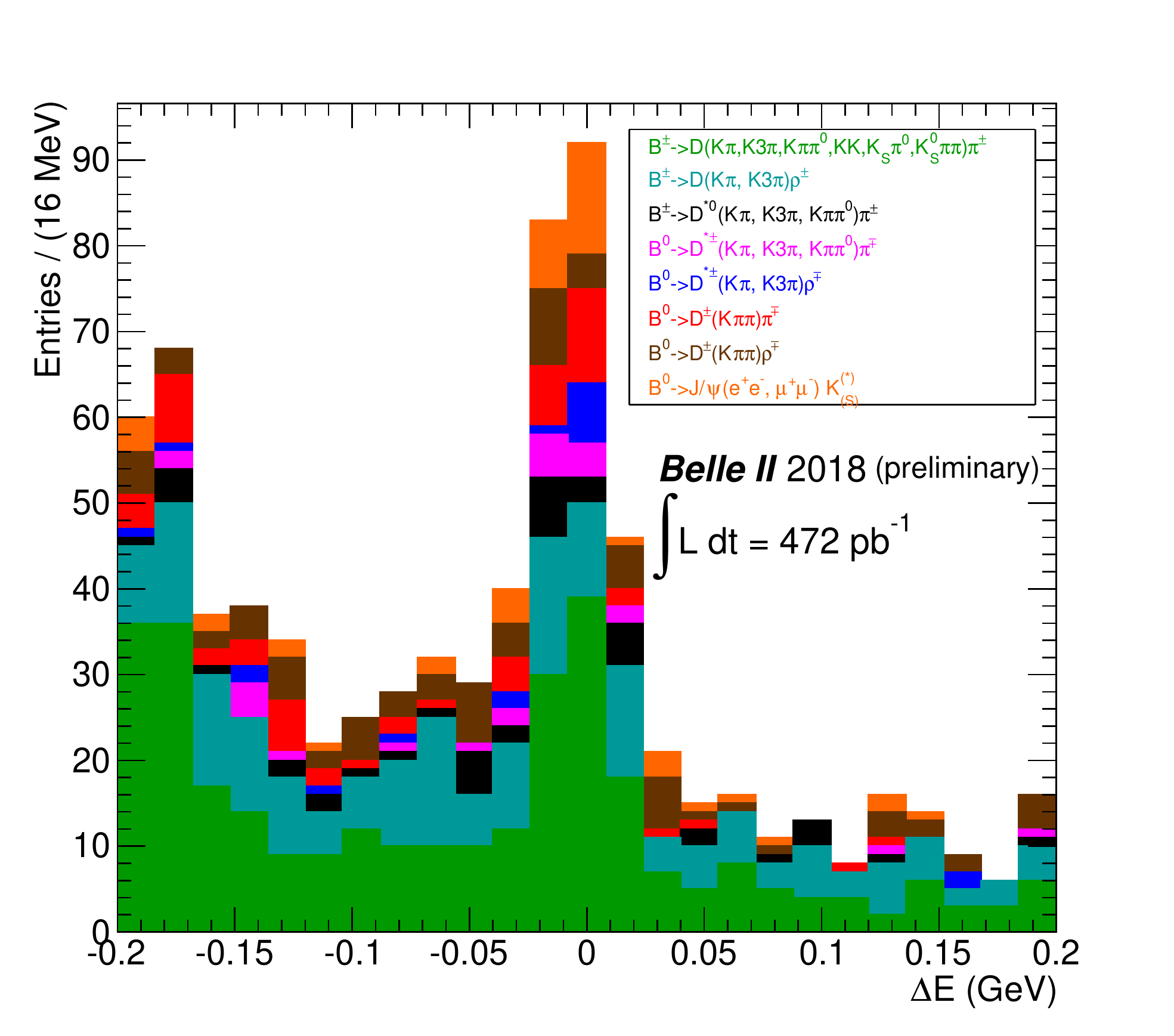} &
\includegraphics[width=0.5\columnwidth]{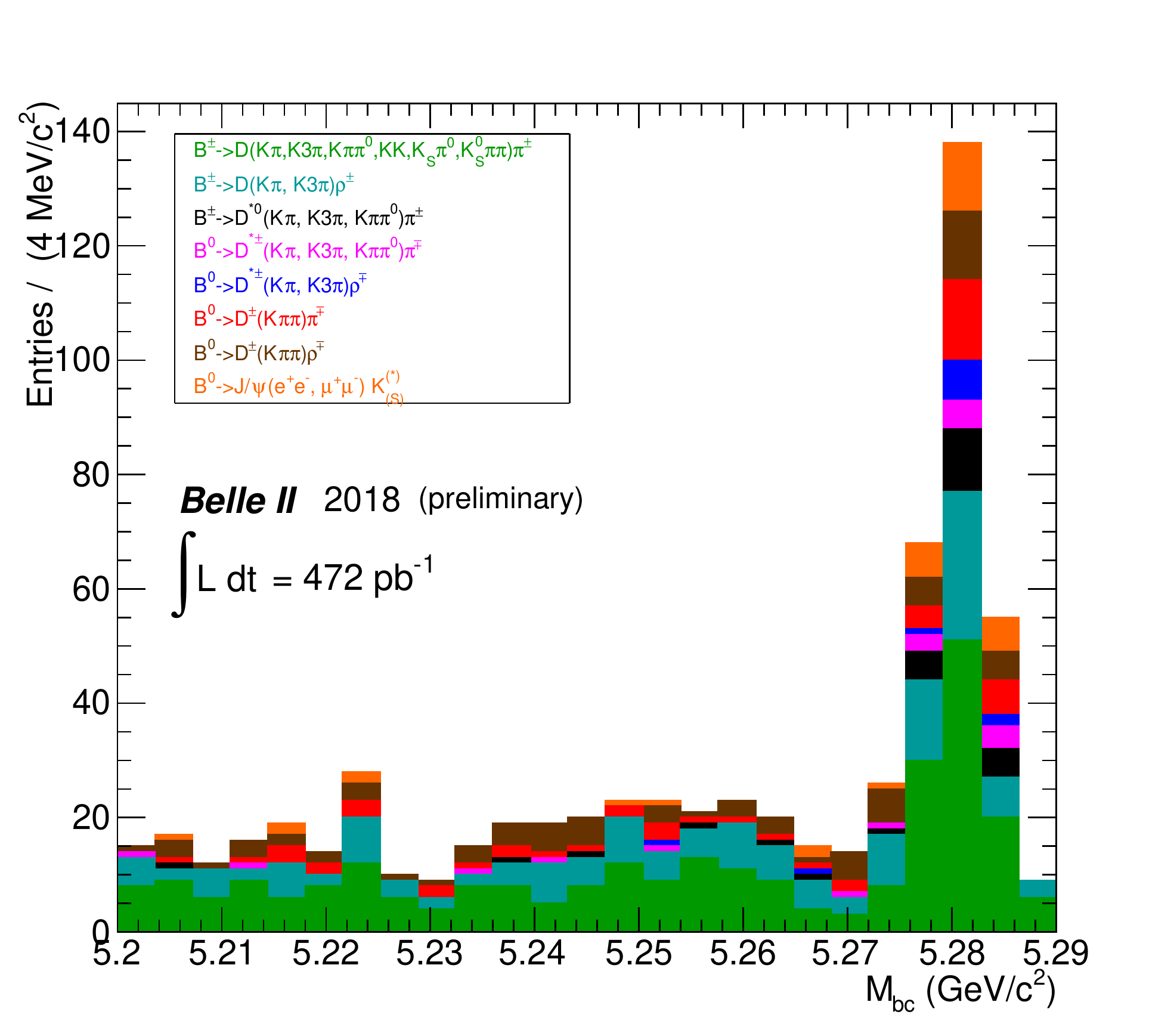} \\
\end{tabular}
\caption{$\Delta E$ (left) and $M_{bc}$ (right) distributions for various $B$ decay modes with phase~II Belle~II data.}\label{Fig:B}
\end{figure}

The $B$ mesons are analysed by defining two kinematic variables the energy difference $\Delta E$ and the beam-constrained mass $M_{\rm bc}$ as $\Delta E = E_{B} - E_{\rm beam}$ and $M_{\rm bc} = c^{-2}\sqrt{E_{\rm beam}^{2}-|\vec{\mathbf{p}}_{B}|^2c^{2}}$, where $E_{B}$ $(\vec{\mathbf{p}}_{B})$ is the energy (momentum) of the $B$ candidate and $E_{\rm beam}$ is the beam energy in the centre-of-mass frame. There are around 245 $B$ candidates observed in the phase~II data from different final states and their $\Delta E$ and $M_{\rm bc}$ distributions are shown in Fig.~\ref{Fig:B}.

%The $B^+\to D^{(*)}\pi^+$ decays are used as an important calibration samples for $\phi_3$ determination. The decays $B^0 \to D^{(*)+}\pi^-$~\cite{Dstpi_tdcpv} and $B^0 \to D^{(*)+}\rho^-$~\cite{Dstrho_tdcpv} are used to extract $\phi_3$ via time-dependent $CP$-violation measurements. The CKM angle $\phi_1$ is determined from $B^0 \to J/\psi K_{\rm S}^0$ decays. So these $B$ rediscoveries illustrate the good prospects for measurements of various CKM parameters at Belle~II.

\section{$\phi_3$-sensitivity at Belle~II}
\label{Sec:phi3}

\begin{figure}[ht!]
\centering
\includegraphics[width=0.55\columnwidth]{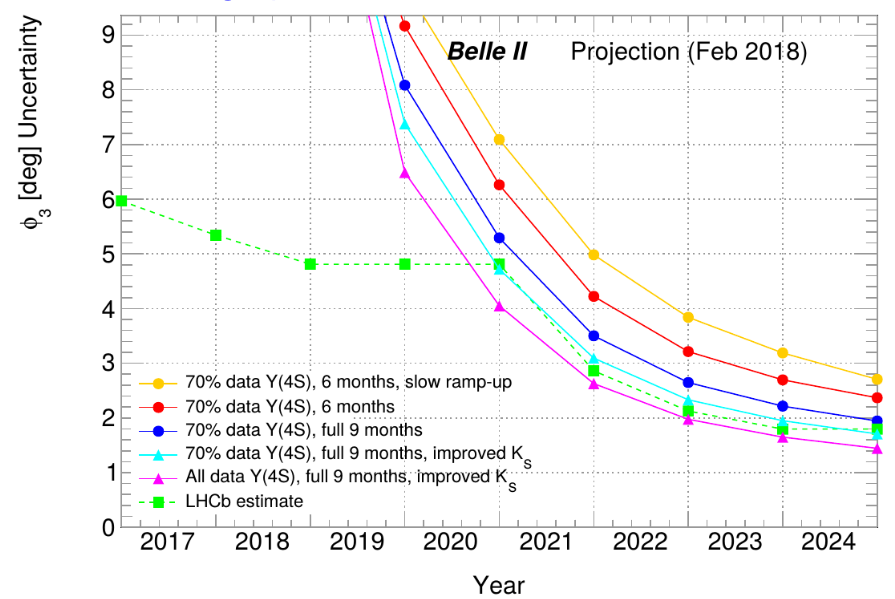}
\caption{Projected sensitivity of $\phi_3$ at Belle~II.}\label{Fig:sen}
\end{figure}

The decays $B^{\pm}\to D(K_{\rm S}^0\pi^+\pi^-)K^{\pm}$ is considered the golden mode to measure $\phi_3$ at Belle~II. The model-independent GGSZ method has been successful in determining $\phi_3$ precisely from this decay mode. Belle~II simulations show that the uncertainty of this $\phi_3$ measurement can be brought down up to 3$^{\circ}$, provided the $D$ strong-phase difference parameters $c_i$  and $s_i$ are measured from the 10~fb$^{-1}$ of data from BESIII~\cite{b2tip}. The sensitivity of $B^{\pm}\to D(K_{\rm S}^0\pi^+\pi^-\pi^0)K^{\pm}$ has been estimated by assuming that the product of reconstruction efficiency and branching fraction is similar to that of $B^{\pm}\to D(K_{\rm S}^0\pi^+\pi^-)K^{\pm}$. This decay mode is expected to provide a sensitivity of 4.4$^{\circ}$ on $\phi_3$~\cite{Resmi}. The GLW modes $B\to D^{(*)}K$ also has significant impact on the projected uncertainty. The expected $\phi_3$-sensitivity at Belle~II as a function of time is shown in Fig.~\ref{Fig:sen}. 

\section{Summary}
\label{Sec:sum}

The precise measurement of $\phi_3$ is important to make precision tests of the standard model description of $CP$ violation. It is important to add more $D$ final states to reduce the statistical uncertainty on the measurement of $\phi_3$. A combined sensitivity of 1.6$^{\circ}$ is expected with the full 50~ab$^{-1}$ of data from Belle~II~\cite{b2tip}. The $D$ decay inputs from BESIII become imperative for achieving this precision. The rediscoveries in the first data from Belle~II shows good prospects for the decay modes involved in CKM measurements.

\end{document}